\documentclass[aps,prd,twocolumn,amssymb,eqsecnum,showpacs,showkeyes,secnumarabic,graphics,floatfix,nofootinbib,tightenlines,longbibliography,superscriptaddress]{revtex4-1}

\usepackage[pdftex]{graphicx,color}
\usepackage{bm}
\usepackage{amsmath,amssymb}
\usepackage[T1]{fontenc}
\usepackage[utf8]{inputenc}
\usepackage[english]{babel}
\usepackage{lmodern}
\usepackage{longtable}
\usepackage{wasysym}
\usepackage[dvipsnames]{xcolor}
\usepackage[breaklinks=true,colorlinks=true,
linkcolor=blue,urlcolor=Blue,citecolor=MidnightBlue,
bookmarks=true,bookmarksopenlevel=2]{hyperref}

\definecolor{valecol}{rgb}{0,0.5, 1.}

\definecolor{leacol}{rgb}{1.,0.5, 0}

\def \om    {\Omega}

\def \om0m {\Omega_{0\rm m}}

\newcommand{\newc}{\newcommand}
\newc{\D}{\partial}
\newc{\ie}{{\it i.e.} }
\newc{\eg}{{\it e.g.} }
\newc{\etc}{{\it etc.} }
\newc{\etal}{{\it et al.}}
\newc{\lcdm}{$\Lambda$CDM }
\newc{\lcdmnospace}{$\Lambda$CDM}
\newc{\wcdm}{$w$CDM }
\newc{\plcdm}{Planck/$\Lambda$CDM }
\newc{\plcdmnospace}{Planck/$\Lambda$CDM}
\newc{\wlcdm}{WMAP7/$\Lambda$CDM }
\newc{\wlcdmnospace}{WMAP7/$\Lambda$CDM}

\newc{\ra}{\Rightarrow}
\newc{\fs}{$f\sigma_8$}
\newc{\fsz}{$f\sigma_8(z)$}

\newc{\bea}{\begin{eqnarray*}}
\newc{\eea}{\end{eqnarray*}}
\newc{\be}{\begin{equation}}
\newc{\ee}{\end{equation}}
\newc{\ba}{\begin{eqnarray}}
\newc{\ea}{\end{eqnarray}}

\begin{document}

\title{Gravitational transitions via the explicitly broken symmetron screening mechanism}

\author{Leandros Perivolaropoulos}\email{leandros@uoi.gr}
\affiliation{Department of Physics, University of Ioannina, GR-45110, Ioannina, Greece}
\author{Foteini Skara}\email{f.skara@uoi.gr}
\affiliation{Department of Physics, University of Ioannina, GR-45110, Ioannina, Greece}

\date{\today}

\begin{abstract}
We generalize the symmetron screening mechanism by allowing for an explicit symmetry breaking of the symmetron $\phi^4$ potential. A coupling to matter of the form $A(\phi)=1+\frac{\phi^2}{M^2}$ 
leads to an explicitly broken symmetry with effective potential $V_{eff}(\phi)=-\mu^2 (1-\frac{\rho}{\mu^2 M^2})\phi^2 +\frac{\lambda}{2}\phi^4 + 2 \varepsilon \phi^3+\frac{\lambda}{2}\eta^4$. Due to the explicit symmetry breaking induced by the cubic term we call this field the 'asymmetron'. For large matter density $\rho>\rho_*\equiv \mu^2M^2+\frac{9}{4}\varepsilon\eta M^2$ the effective potential has a single minimum at $\phi=0$ leading to restoration of General Relativity (GR) as in the usual symmetron screening mechanism. For low matter density however, there is a false vacuum and a single true vacuum due to the explicit symmetry breaking. This is expected to lead to an unstable network of domain walls with slightly different value of the gravitational constant $G$ on each side of the wall. This network would be in constant interaction with matter overdensities and would lead to interesting observational signatures which could be detected as gravitational and expansion rate transitions in redshift space. Such a gravitational transition has been recently proposed for the resolution of the Hubble tension.
\end{abstract}
\maketitle

\section{Introduction}
It has recently been pointed out \cite{Marra:2021fvf,Alestas:2020zol,pres1} that a fundamental physics phase transition taking place at a redshift $z_t\lesssim 0.01$ and leading to a sudden increase of the type Ia supernovae (SnIa) absolute magnitude $M$ by about $\Delta M \simeq 0.2$ for $z<z_t$ \cite{Marra:2021fvf} can lead to a resolution of the Hubble tension \cite{DiValentino:2020zio} between
the Planck estimate \cite{Planck:2018vyg} and the SH$0$ES collaboration measurements  \cite{Riess:2021jrx} which is currently at the $5\sigma$ level  (see Refs. \cite{Abdalla:2022yfr,Perivolaropoulos:2021jda,DiValentino:2021izs,Verde:2019ivm} for recent reviews and Refs. \cite{Alestas:2021xes,Alestas:2021luu,Alestas:2020mvb,Kazantzidis:2020tko,DiValentino:2020vnx,Theodoropoulos:2021hkk} for recent studies of the ultra-late transition approach to the Hubble tension). Under simple assumptions about the connection of the SnIa absolute magnitude with the effective gravitational constant $G_{eff}$ \cite{Amendola:1999vu,Gaztanaga:2001fh,Wright:2017rsu} , this transition could be induced by a gravitational transition increasing the value of the gravitational constant up to about $10\%$ for $z<z_t$. If such a transition were to imply weaker gravity \cite{Amendola:1999vu,Gaztanaga:2001fh} in the past it could also play an important role in the resolution of another tension of the standard \lcdm model known as the $\sigma_8$ or 'growth' tension \cite{DiValentino:2020vvd,Macaulay:2013swa,Hildebrandt:2016iqg,Joudaki:2017zdt,Nesseris:2017vor,Kazantzidis:2018rnb,Skara:2019usd,Kazantzidis:2019nuh,Perivolaropoulos:2019vkb,Kazantzidis:2020xta}.

In view of the effectiveness of such a transition in the resolution of the Hubble and growth tensions, the following questions emerge
\begin{itemize}
    \item Is such a transition consistent with current observational and experimental constraints on the evolution of $G_{eff}$?
    \item Are there any hints in observational data for such a transition?
    \item Are there theoretical models \cite{Tsujikawa:2015mga,Gannouji:2018ncm,Gannouji:2020ylf} that can generically predict such a transition at the spatial or temporal level at $z_t \lesssim 0.01$?
\end{itemize}
The answer to the first question is positive. In fact current constraints on the evolution of $G_{eff}$ strongly constrain its time derivative at present and at specific times and distances in the past. However a abrupt shift of $G_{eff}$ is weakly constrained and the current bounds allow an abrupt change of $G_{eff}$ by up to about $5-10\%$ at some cosmological time in the past between the present time and the time of nucleosynthesis. 

The answer to the second question is also positive. Hints for such a transition in the values of dynamical parameters connected to the gravitational constant have recently been pointed out in Cepheid SnIa calibrator data \cite{Perivolaropoulos:2021bds,Mortsell:2021nzg}, in Tully-Fisher data \cite{Alestas:2021nmi} and in solar system history data \cite{Perivolaropoulos:2022vql} which indicate an increase of the rate of impactors on the Moon and Earth surfaces by about a factor of 2-3 during the past 100Myrs which correspond to $z<0.008$ \cite{1998JRASC..92..297S,1995hdca.book.....G,1997JGR...102.9231M,2001JGR...10632847G,Ward2007TerrestrialCC,2019Sci...365.9895M,Bottke}. Such a transition is also consistent with low redshift galaxy surveys data \cite{Alestas:2022xxm}.

The answer to the third question may be approached at both the temporal and the spatial level. In the context of a temporal transition a nonminimal scalar field could be initially trapped either due to cosmic friction or due to a local minimum of a time-dependent potential and globally shift to a new minimum of the effective potential at $z_t$ via a classical evolution of the potential which may be coupled to the matter density or via the reduction of the cosmic friction. An alternative scenario leading to a gravitational transition could include a pressure non-crushing cosmological singularity in the recent past \cite{Odintsov:2022eqm}.

In the context of a tunneling first order phase transition of spatial character we, as observers, may be located in a true or false vacuum bubble with scale of about $20-40Mpc$ corresponding to $z<0.01$ where the value of $G_{eff}$ is up to about $10\%$ higher than the value of $G_{eff}$ of the other vacuum of a non-minimally coupled scalar field. 

Alternatively, a mechanism involving a transition with spatial character  by a  purely classical evolution may be realized in the context of a symmetron field used as a screening mechanism  of modified gravity theories. Based in part on earlier work \cite{Olive:2007aj,Pietroni:2005pv} the authors of Ref. \cite{Hinterbichler:2010es} proposed the symmetron screening mechanism  with a specific form of the scalar-gravity coupling where the coupling strength is the density-dependent quantity. The scalar field is decoupled from matter and screened when the matter density is  sufficiently high, while in regions of low density the scalar field is coupled to matter with a long-range mediated force of gravitational strength  \cite{Hinterbichler:2010es,Hinterbichler:2011ca} (see also Refs. \cite{Davis:2011pj,Gronke:2013mea,Gronke:2014gaa,Llinares:2014zxa,Gronke:2015ama,Voivodic:2016kog,Hammami:2015ela} and the next Section for details).

At early times when the mean density of the universe is $\rho>\rho_*$ (where $\rho_*$ is a critical density), the minimum of the effective potential everywhere is at $\phi=0$ and GR is applicable. As the mean density drops below $\rho_*$ the symmetry is spontaneously broken and the symmetron field relaxes at one of the minima, the potential develops in low density regions while in regions where density perturbations have grown to densities above $\rho_*$ the field remains at the symmetric vacuum $\phi=0$. Low density regions where the field has relaxed in different vacua are separated by symmetron domain walls\footnote{A domain wall is a type of two dimensional (sheet-like) topological defect  (solitonic configurations of field) in three spatial dimensions that occurs whenever a discrete symmetry of the potential is spontaneously broken \cite{Kibble:1976sj,Zeldovich:1974uw,Kibble:1982dd,Vilenkin:1984ib,Vachaspati:1984dz,Hindmarsh:1994re,Vilenkin:2000jqa,Vachaspati:2006zz,Manton:2004tk,Perivolaropoulos:2018cgr,Alestas:2019wtw}. It separates neighboring spatially domains where the field is in different vacua.} where the field by continuity goes through the local maximum of the potential $\phi=0$.  Due to the $\mathbb{Z}_2$ symmetry of the potential $\phi^2$ is the same at the two vacua and the corresponding effective gravitational constant $G_{eff}$ in the Jordan frame,  is the same on the two sides of the symmetron wall. Thus in the context of the symmetron domain wall no transition of $G_{eff}$ is expected as the symmetron domain wall is crossed. 

This is not the case if the bare potential includes an explicit $\mathbb{Z}_2$ symmetry breaking term $\varepsilon \phi^3$.  In this case the two local minima of the potential in low density regions are not symmetric ($\phi_+\neq -\phi_-$) and this implies a transition in the value of the Jordan frame gravitational constant as the wall is crossed. In addition the coexistence of a true with a false vacuum implies that the wall network dynamics will involve instabilities and will thus be different from the wall network appearing in the context of symmetric equivalent vacua.

This work focuses on a symmetron mechanism that involves explicit symmetry breaking. For definiteness we call this type of generalized symmetron field the {\it asymmetron}. 

There are at least three main mechanisms that can lead to a gravitational transition observed in the recent cosmological lookback time:
\begin{itemize}
    \item 
Evolving scalar field (extended quintessence) in a scalar tensor sharply varying scalar-tensor potential.
\item
False vacuum decay (first order phase transition) in the context of a scalar-tensor theory.
\item
A network of symmetron domain walls with explicitly broken $Z_2$ symmetry of the effective potential (asymmetron wall network).
\end{itemize}
The present analysis focuses on the third mechanism and aims to provide a better understanding of the scalar field dynamics involved in such a mechanism. The main questions addressed in this context are the following:
\begin{itemize}
\item
How can a gravitational transition be realized in the context of an asymmetron wall network?
\item
What are the properties and evolution of an asymmetron field domain wall in the presence of a spherical matter shell overdensity?
\item
Are there cosmological observations that could be interpreted as results on an existing asymmetron domain wall network?
\end{itemize}

The paper is structured as follows. Section \ref{symmetron} introduces the necessary background and notation of symmetron screening. In Section \ref{enerdyn} we introduce the asymmetron field and the explicit $Z_2$ symmetry breaking associated with it. We also present the energetics and dynamics of spherical symmetron and asymmetron domain walls. Static stable wall solutions in the presence of matter are derived in Section \ref{staticsol}. We also point out that recent cluster profile data may be interpreted as revealing spatial cosmological sectors where distinct properties of gravity are present. We discuss the possible connection of such an effect with the existence of asymmetron domain walls.
Finally in Section \ref{CONCLUSION-DISCUSSION} we conclude, summarise and discuss possible extensions of our analysis.  

In what follows we assume a metric signature $(-,+,+,+)$.

\section{Review of the symmetron screening}
\label{symmetron}
In the context of the symmetron mechanism\footnote{For reviews of modified gravity theories with screening mechanisms, such as the Vainshtein  \cite{Vainshtein:1972sx,Arkani-Hamed:2002bjr,Deffayet:2001uk} and the chameleon \cite{Khoury:2003aq,Khoury:2003rn,Gubser:2004uf,Brax:2004qh,Brax:2004px,Upadhye:2006vi,Mota:2006ed,Mota:2006fz,Brax:2008hh,Brax:2010kv} models see in Refs. \cite{Khoury:2010xi,Burrage:2017qrf,Sakstein:2013pda,Brax:2012gr,Jain:2010ka,Davis:2011qf,Hui:2009kc,Burrage:2016bwy,Joyce:2014kja,Brax:2021wcv,Baker:2019gxo,Sakstein:2018fwz}.}, screening is achieved via $\mathbb{Z}_2$ symmetry restoration in regions with matter density larger than a critical density. 

The symmetron model is a special case of a general scalar-tensor theory, thus its action in the Einstein frame (where the scalar field couples non-minimally to matter components and minimally to gravity) is described by  the general scalar-tensor action  \cite{Hinterbichler:2010es,Hinterbichler:2011ca,Davis:2011pj,Gronke:2013mea,Gronke:2014gaa,Llinares:2014zxa}
\begin{align}
S&=\int d^4 x\sqrt{-g}\left[\frac{R}{16\pi G}-\frac{1}{2}\nabla_\mu\phi\nabla^\mu\phi-V(\phi)\right]\nonumber\\
& +S_m\left[\psi_i,{\tilde g}_{\mu \nu}\right]
\label{eq:STtheoriesgeneral}
\end{align}
where $G$ is Newton’s constant as measured locally e.g. in Eotvos-type experiments, $g$ is the determinant of the Einstein frame metric $g_{\mu\nu}$, $R$ is the Ricci scalar, $\phi$  is a scalar field with self-interactions given by the potential $V(\phi)$, $S_m$ is the action for the various matter fields and $\psi_i$ represent these  matter fields which are minimally coupled in the Jordan frame metric ${\tilde g}_{\mu\nu}$\footnote{In the rest of this paper, quantities associated to the Jordan frame metric ${\tilde g}_{\mu\nu}$ will be distinguished by a tilde.}. This is connected to the Einstein frame metric $g_{\mu\nu}$ via a conformal rescaling  \cite{Hinterbichler:2010es,Hinterbichler:2011ca,Davis:2011pj,Gronke:2013mea,Gronke:2014gaa,Llinares:2014zxa}
\begin{equation}
\label{eq:Weylgen}
{\tilde g}_{\mu \nu}=A^2(\phi)g_{\mu \nu}
\end{equation}
The non-minimal coupling to matter is described by the coupling function $A(\phi)$ and leads to deviations from GR. The scalar field couples to the  trace of the energy-momentum tensor and its equation of motion, obtained using standard variational methods, is \cite{Hinterbichler:2010es,Hinterbichler:2011ca}
\begin{equation}
\label{eq:EOMgen}
\Box\phi=\frac{d V(\phi)}{d \phi}-\frac{d A(\phi)}{d \phi}A(\phi)^3\tilde{T}
\end{equation}
where $\tilde{T}$ is the trace $\tilde{T}=\tilde{g}_{\mu\nu}\tilde{T}^{\mu\nu}$ of the Jordan frame energy-momentum tensor 
\begin{equation}
\tilde{T}^{\mu\nu}\equiv \frac{-2}{\sqrt{-\tilde{g}}}\frac{\delta S_m}{\delta \tilde{g}_{\mu\nu}}=A(\phi)^{-6}T^{\mu\nu}
\end{equation}
which is covariantly conserved $\tilde{\nabla}_{\mu}\tilde{T}^{\mu\nu}=0$.

For non-relativistic matter the trace of the Einstein energy-momentum tensor\footnote{Note that in the Einstein frame the density $\rho$ is not conserved but  the ’density’ $ A(\phi)^3\tilde{\rho}$ is conserved \cite{Hinterbichler:2010es,Waterhouse:2006wv} and $\phi$-independent \cite{Hinterbichler:2011ca}. However the coupling function is assumed to be a weak function of $\phi$ ($A(\phi)\approx1$), so that the two densities do not differ from each other significantly  ($\rho\approx A(\phi)^3\tilde{\rho}$). }  is
$T=-\rho\approx-A(\phi)^3\tilde{\rho}=-A(\phi)^3\tilde{T}$, and the scalar field equation of motion (\ref{eq:EOMgen}) takes the form
\begin{equation}
\Box\phi=\frac{d V(\phi)}{d\phi}+\frac{\beta(\phi)\rho}{M_{pl}}=\frac{d V_{eff}}{d\phi}
\label{eomsym}
\end{equation}
where $M_{pl}=(8\pi G)^{-1/2}$ is the reduced Planck mass, $V_{eff}$ is the effective potential\footnote{Note that in the literature the effective potential is often defined as  $V_{eff}(\phi)=V(\phi)+\rho[A(\phi)-1]$ \cite{Brax:2012gr,Gronke:2013mea,Gronke:2014gaa,Sami:2021ufn} or  $V_{eff}(\phi)=V(\phi)+\rho\ln A(\phi)$ \cite{Burrage:2017qrf}.} \cite{Hinterbichler:2010es,Hinterbichler:2011ca}
\begin{equation}
\label{eq:veff_gen}
V_{eff}(\phi)=V(\phi)+\rho A(\phi)
\end{equation}
and the $\beta$ is the coupling between the
scalar field and matter
\begin{equation}
\beta(\phi)=M_{pl}\frac{d A(\phi)}{d \phi}
\label{beta}
\end{equation}
This coupling characterises the strength of the scalar  fifth force which, in the
nonrelativistic limit, is given by \cite{Waterhouse:2006wv,Davis:2011pj,Winther:2011qb}
\be
\vec{F}_{\phi} =\frac{\beta(\phi)}{M_{pl}}\vec{\nabla} \phi
\label{force}
\ee
This scalar fifth force is an additional contribution to the (Newtonian) gravitational force $F_N$. 

The interaction potential and the coupling function are chosen to be of the spontaneous symmetry breaking form \cite{Hinterbichler:2010es,Hinterbichler:2011ca,Davis:2011pj}
\begin{equation}
V(\phi)=\frac{\lambda}{4}(\phi^2 -\eta^2)^2 
\label{potsym}
\end{equation}
\be
A(\phi)=1+\frac{\phi^2}{2M^2}+\mathcal{O}(\frac{\phi^4}{M^4})
\label{couplf}
\ee
where $M$ is the mass scale of symmetron field coupling to the matter density. It gives the strength of the interaction with the matter fields. The parameter $\lambda$ is a positive dimensionless coupling securing that the energy of the $\phi^4$ model \cite{Dashen:1974cj,Polyakov:1974ek} is bounded from below \cite{Manton:2004tk}). Also $\eta=\phi_0=\phi(\rho=0)$ is the expectation value of the scalar field at zero matter density. For the field range  $(\frac{\phi}{M})^2\ll 1$ the higher order correction terms of the coupling function can be consistently neglected \cite{Hinterbichler:2011ca,Davis:2011pj}.

The effective potential is
\begin{equation}
V_{eff}(\phi)=-\frac{1}{2}\mu^2\left(1-\frac{\rho}{\mu^2M^2}\right)\phi^2+\frac{\lambda}{4}\phi^4 + \frac{\lambda\eta^4}{4}
\label{veffsym}
\end{equation}
where $\mu^2\equiv \lambda \eta^2$. 

The effective potential is invariant with respect  to the $\mathbb{Z}_2$ symmetry (reflection symmetry) transformation $\phi\rightarrow-\phi$ (as are $V(\phi)$ and $A(\phi)$ individually). The coefficient of the quadratic term (effective mass) changes sign at a critical density 
\begin{equation}
\rho_* \equiv \mu^2 M^2
\end{equation}
For density smaller than the critical density ($\rho<\rho_*$) the effective mass is negative, the $\mathbb{Z}_2$ symmetry is spontaneously broken and the effective potential has two nonzero degenerate minima  located at
\begin{equation}
\phi_{\pm}=\pm\eta\sqrt{1-\frac{\rho}{\rho_*}}
\label{phimin}
\end{equation}
leading to two degenerate vacua. Note that if $\rho\ll\rho_*$ then the vacua correspond to  $\phi_{\pm}\approx\pm\eta=\pm\frac{\mu}{\sqrt{\lambda}}$. 

For background density larger than the critical density ($\rho>\rho_*$) the symmetry gets restored (symmetric phase) and the effective potential has a unique global minimum at the origin ($\phi= 0$) about which it is symmetric. 

From Eqs. (\ref{beta}) and (\ref{couplf}) the coupling to matter at the  minima of the effective potential  is given by
\be   
\beta(\phi_{\pm})=\frac{M_{pl}\phi_{\pm}}{M^2}= 
\begin{cases}
    0&\rho>\rho_*\\
    \pm \beta_0\sqrt{1-\frac{\rho}{\rho_*}}              & \rho<\rho_*
\end{cases}
\ee
where $\beta_0\equiv\frac{M_{pl}\eta}{M^2}$ is the coupling at zero matter density (vacuum). Clearly, the strength of the coupling to matter depends on the background density. Thus in high density regions the field does not couple to matter and the fifth force in Eq. (\ref{force}) is suppressed  while in regions of low density the field couples to matter and mediates a force.

Using Eq. (\ref{veffsym}) we have for the effective mass of the symmetron field
\begin{align}
m_{eff}^2\equiv &\left.\frac{d^2 V_{eff}}{d\phi^2}\right|_{min}=\left(\frac{\rho}{\rho_*}-1\right)\mu^2+3\lambda \phi_{\pm}^2\Rightarrow \nonumber\\
&m_{eff}^2=2\mu^2\left(1-\frac{\rho}{\rho_*}\right)
\end{align}
and the range (Compton wavelength) of the field in density regions with $\rho < \rho_*$ is 
\be
l_{\phi}= \frac{1}{m_{eff}}=\frac{1}{\sqrt{2}\mu}\left(1-\frac{\rho}{\rho_*}\right)^{-1/2}
\label{range}
\ee

The spontaneous symmetry breaking phase can lead to the formation of a domain wall network via the Kibble mechanism. These walls are attracted to high density regions (see in Refs. \cite{Llinares:2014zxa,Pearson:2014vqa,Peyravi:2016cnk} for numerical studies of properties and dynamics of domain walls in the symmetron model). The physical origin of this interaction is described in the next section. The profile of such a static domain wall with boundary conditions $\phi(x\rightarrow \pm \infty) =\phi_{\pm}$, is obtained by solving Eq. (\ref{eomsym}) and is of the form
\be
\phi(x)=\eta\sqrt{1-\frac{\rho}{\rho_*}}\tanh\left[\sqrt{\frac{\lambda}{2}}\eta\sqrt{1-\frac{\rho}{\rho_*}}x\right]
\ee
Its width is 
\be
\delta=\frac{1}{\mu}\left(1-\frac{\rho}{\rho_*}\right)^{-1/2}=\sqrt{2}l_{\phi}
\ee
A slowly evolving wall network may be interpreted as a fluid with equation of state parameter \cite{Vachaspati:2006zz}
\be
w_w=\frac{p_w}{\rho_w}=-\frac{2}{3}
\ee
and density parameter \cite{Llinares:2014zxa}
\be
\Omega_w\equiv\frac{\rho_w}{\rho_c}=\frac{\sigma}{3H^2M_{pl}^2a\,d}
\ee
where $a$ is the scale factor and $d$ is the comoving distance between the walls\footnote{Assuming parallel domain walls separated by physical distance  $a\,d$ which grows in proportion with the scale factor.}, $\sigma \equiv \rho_w \,a\,d$  is the surface energy density (energy per unit area or tension) of the wall (with $\sigma=\frac{4}{3}\sqrt{\frac{\lambda}{2}}\eta^3$ for $\rho=0$ \cite{Vachaspati:2006zz}),  $\rho_c$ is the critical density of the universe and $H=\dot{a}/a$ is the Hubble parameter.

In theories where the phase transition takes place in the recent past (around the onset of cosmic acceleration) the scale factor at the time of the symmetry breaking is given by \cite{Davis:2011pj}
\be
a_*^3=\frac{\rho_0}{\rho_*}=\frac{3\Omega_{0m}H_0^2M_{pl}^2}{\mu^2 M^2}
\ee
where $\rho_0=\rho(a=1)$ and $\Omega_{0m}=\Omega_m(a=1)$ are the matter  density and the corresponding density parameter  in the universe today respectively while $H_0$ is the Hubble constant. This equation fixes $\mu$ in terms of $M$ and hence combining with the Eq. (\ref{range}) we obtain for redshifts $z<z_*$ (with $z_*=\frac{1}{a_*} -1$) in low density regions ($\rho \ll \rho_*$)
\be
l_{\phi}^2\simeq\frac{M^2}{6\,\Omega_{0m}M_{pl}^2 H_0^2}\frac{1}{(1+z_*)^3}
\ee
As shown in Eq. (\ref{range}), in density regions with $\rho< \rho_*$  there is a dependence of the symmetron range on the background matter density and hence the redshift. The range decreases as the redshift at the time of symmetry breaking increases. For a range $M \lesssim 10^{-3}M_{pl}$, the range of the scalar field force becomes $l_{\phi} \lesssim 1 Mpc$ \cite{Khoury:2010xi,Davis:2011pj,Sakstein:2014jrq}. The value $1 Mpc$ corresponds intergalactic distance in clusters and therefore dynamical observational cosmological effects are anticipated for this range.

The background cosmology, the evolution of  perturbations and large-scale structure in the context of the symmetron model have been investigated in \cite{Hinterbichler:2011ca,Davis:2011pj,Brax:2011pk,Clampitt:2011mx,Llinares:2012ds,Taddei:2013bsk}. Before the time of the symmetry breaking ($t<t_*$) we have $\phi\approx 0$ and the effective gravitational constant $G_{eff}=G$. While after the symmetry breaking ($t>t_*$) the field approaches the minima  $\phi_{\pm}=\pm \eta$ in low density regions and the effective gravitational constant is (see in Ref. \cite{Davis:2011pj} for details) 
\be
   G_{eff}= 
\begin{cases}
    G& a/k\gg l_{\phi}\\
    G\left(1+2\beta_0^2\right)      & a/k\ll l_{\phi}
\end{cases}
\ee
The implementation of N-body simulations constitutes a useful tool for cosmological studies and for observational predictions of the symmetron screening mechanism  \cite{Davis:2011pj,Winther:2011qb,Brax:2012nk,Llinares:2013qbh,Llinares:2013jua,Llinares:2013jza,Gronke:2013mea,Hagala:2015paa}. 

\section{Asymmetron Domain Walls}
\label{enerdyn}

In this Section we generalize the symmetron mechanism by allowing for an explicit symmetry $Z_2$ breaking of the symmetron potential (\ref{potsym}). The explicit symmetry breaking is induced by the inclusion of a cubic term $\varepsilon \phi^3$ in the potential. In this case the two local minima of the effective potential in low density regions are not symmetric ($\phi_+\neq -\phi_-$). We call this generalized symmetron field the {\it asymmetron}. 

The explicit symmetry breaking can create domain walls which interpolate between spatial regions with the vacuum values $\phi_+$ and $\phi_-$. Also the coexistence of a true with a false vacuum implies that the wall network dynamics will involve instabilities in contrast to the wall network appearing in the case of symmetron model equivalent vacua $\phi_+= |\phi_-|$. In addition it can lead to a transition in the value of gravitational constant  $G$ as the wall is crossed.  Before the time of the symmetry breaking ($t<t_*$) we have $\phi\approx 0$ and the effective gravitational constant is $G_{eff}=G$ as in the case of symmetron field. After the symmetry breaking ($t>t_*$) the field approaches different minima  $\phi_+=\eta_1$ and $\phi_-=\eta_2$ in different domains. The effective gravitational constant is
\be
   G_{eff}= 
\begin{cases}
    G& a/k\gg l_{\phi}\\
    G\left(1+2\beta_{0i}^2\right)       & a/k\ll l_{\phi}
\end{cases}
\ee
where $\beta_{0i}\equiv\frac{M_{pl}\eta_i}{M^2}$ (with $i=1,2$) are the coupling at the true and false vacua. Thus, low density regions in different domains would have different values of gravitational constant and thus different expansion rates since $H^2 \sim G_{eff}$.
\begin{figure}
\centering
\includegraphics[width = \columnwidth]{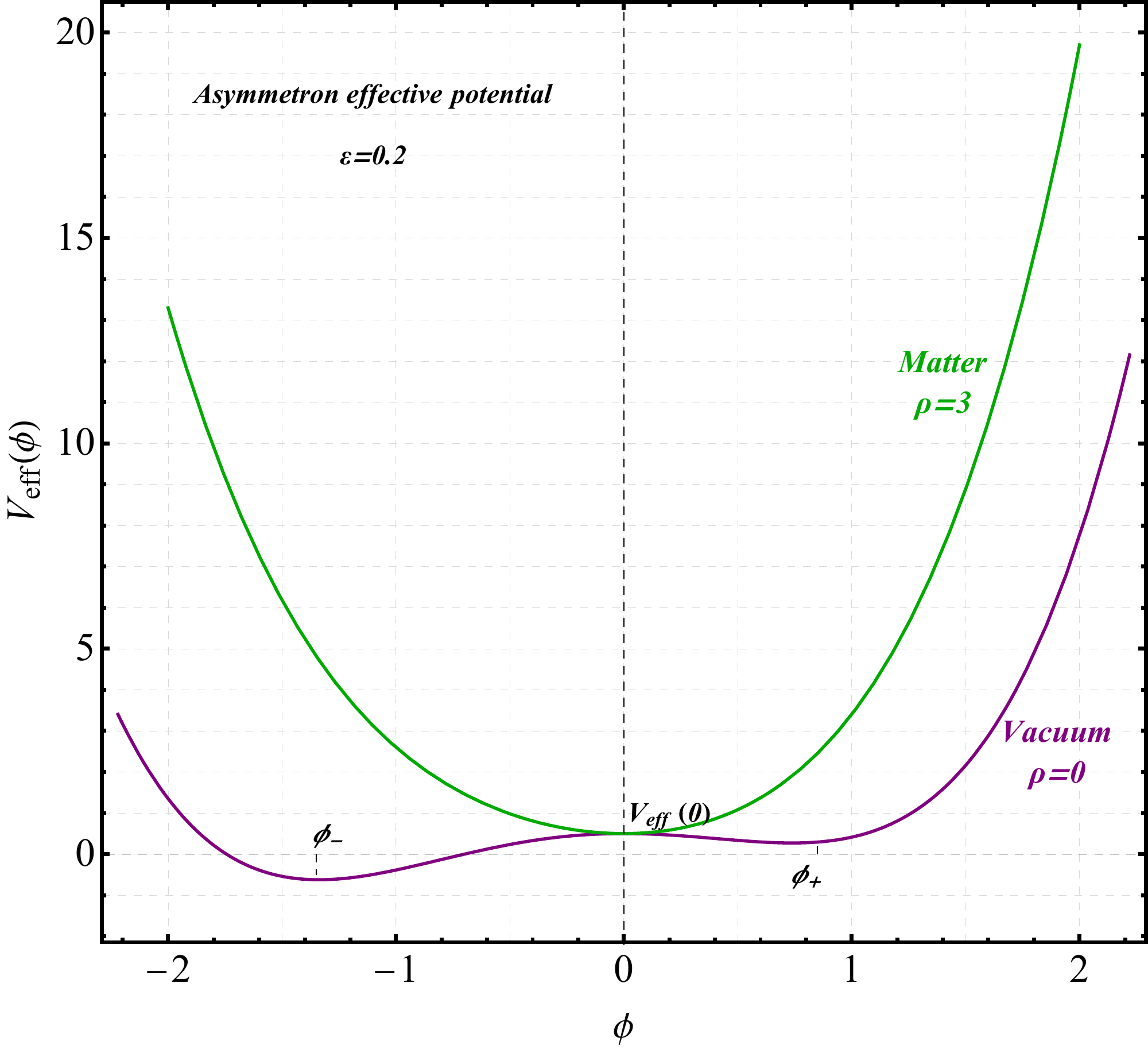}
\caption{
Schematic plots of the asymmetron effective potential Eq. (\ref{poteff}) in vacuum (purple) and in high density (green) cosmological regions. Notice the asymmetric form of the effective potential in which the degeneracy of the vacua is slightly broken. However in the presence of sufficiently high density, a single minimum at $\phi=0$ restores GR as in the symmetron case. }
\label{figpot}
\end{figure}

\subsection{Dynamical equations and energetics of spherical asymmetron configurations}
The action describing the dynamics of the symmetron scalar field may be written as\footnote{We have multiplied by a factor of 2 the usual form of the action to avoid the factor of $\frac{1}{2}$ in the kinetic term.}
\be
S=\int d^4 x\sqrt{-g}\left[g^{\mu \nu}\partial_{\mu}\phi \partial_{\nu}\phi
-V(\phi)\right]
\label{action}
\ee
The dynamical equation for a spherically symmetric field configuration in flat space is
\be
r^2\ddot{\phi}-\frac{\partial}{\partial r}r^2\frac{\partial\phi}{\partial r}=-\frac{1}{2}\frac{d V}{d\phi}r^2
\label{eom}
\ee
where the dot denotes differentiation with respect to cosmic time $t$.

The corresponding energy is
\be
E=4\pi\int_0^{\infty}r^2\left[\left(\frac{d\phi}{dr}\right)^2+V(\phi)\right]dr
\label{energy}
\ee
We now assume a $\phi^4$ potential which includes an explicit $\mathbb{Z}_2$ symmetry breaking term $\varepsilon \phi^3$
\be
V(\phi)=\frac{\lambda}{2}(\phi^2-\eta^2)^2+2\varepsilon \phi^3
\ee
with a coupling to matter
\be
A(\phi)=1+\frac{\phi^2}{M^2}
\ee
such that the effective potential is
\be
V_{eff}(\phi)=-\mu^2 (1-\frac{\rho}{\mu^2 M^2})\phi^2 +\frac{\lambda}{2}\phi^4 + 2 \varepsilon \phi^3+\frac{\lambda}{2}\eta^4
\ee
where $\varepsilon$ is a  parameter. 

By defining the effective rescaled potential
$\bar{V}_{eff}(\bar{\phi})\equiv V_{eff}(\phi)/\lambda\eta^4$ we obtain
\be
\bar{V}_{eff}(\bar{\phi})=-\left(1-\bar{\rho}\right)\bar{\phi}^2+\frac{1}{2}\bar{\phi}^4+\frac{1}{2}+2\bar{\varepsilon}\bar{\phi}^3
\label{poteff}
\ee
where the rescaled dimensionless
quantities are 
\be
\bar{\phi}\equiv\frac{\phi}{\eta},\,\,\bar{\rho}\equiv\frac{\rho}{\lambda\eta^2 M^2}\equiv\frac{\rho}{m^2 M^2},\,\,\bar{\varepsilon}\equiv \frac{\varepsilon}{\lambda\eta}
\ee
We set also
\be
\bar{r}\equiv r\sqrt{\lambda}\eta,\,\, \bar{E}\equiv\frac{E\sqrt{\lambda}}{4\pi\eta}
\ee
and by taking into account the above redefinitions, we can rewrite the dynamical equation (\ref{eom}) as
\be
\bar{r}^2\ddot{\bar{\phi}}-\frac{\partial}{\partial \bar{r}}\bar{r}^2\frac{\partial\bar{\phi}}{\partial \bar{r}}=-\frac{1}{2}\frac{d \bar{V}_{eff}}{d\phi}(\bar{\phi})\bar{r}^2
\label{eomres}
\ee
and the corresponding energy Eq. (\ref{energy}) as
\be
\bar{E}=\int_0^{\infty} \,\bar{r}^2\left[\left(\frac{d\bar{\phi}}{d\bar{r}}\right)^2+\bar{V}_{eff}(\bar{\phi})\right]d\bar{r}
\label{energyres}
\ee
We omit bar from now on and work with dimensionless quantities.
\begin{figure}
\centering
\includegraphics[width = \columnwidth]{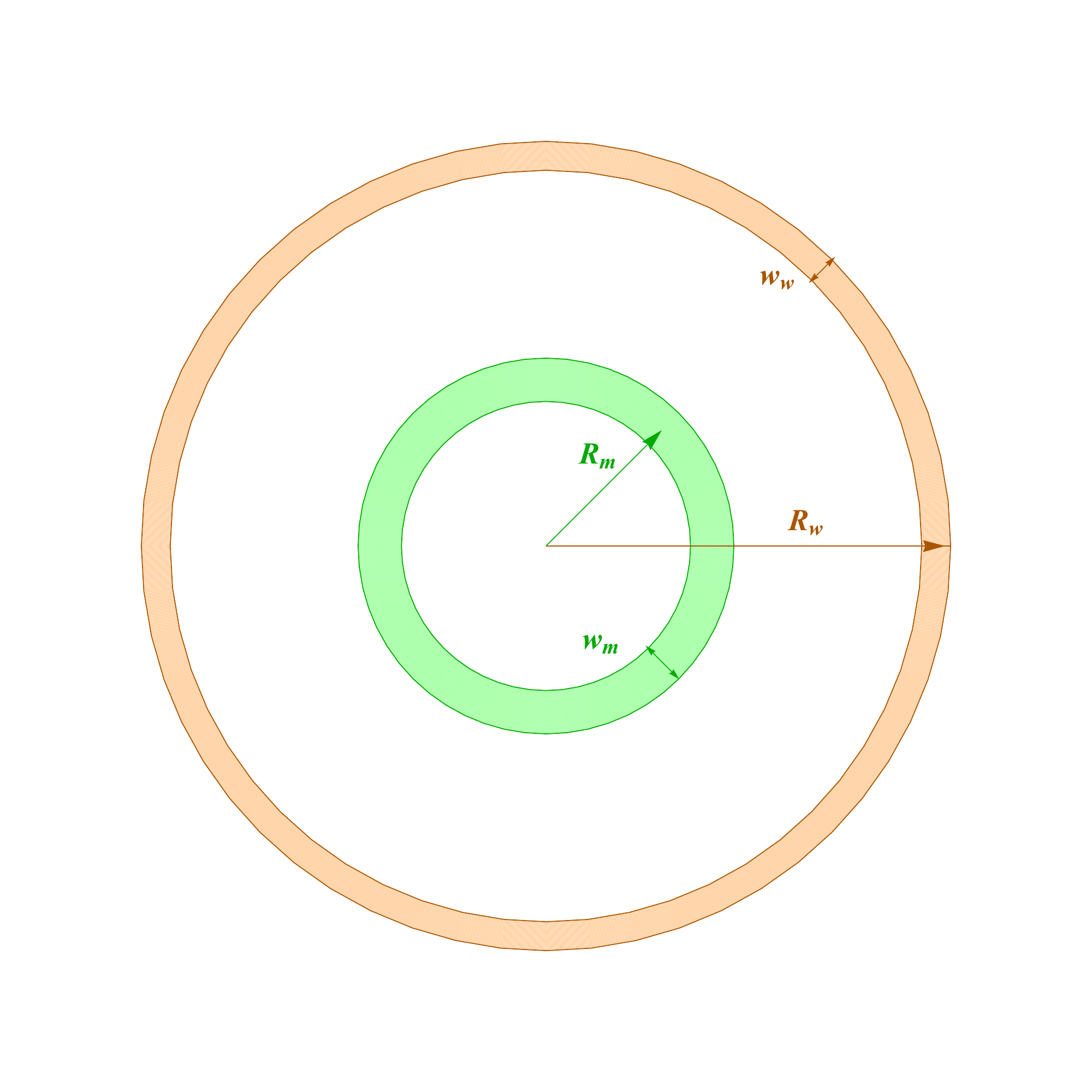}
\caption{The geometry of the spherical domain wall in the presence of spherical matter shell.}
\label{figshells}
\end{figure} 

The two vacuum values (true and false) of $\phi$, given by the equation
\be
\left. \frac{d V_{eff}}{d\phi}\right|_{\phi_{\pm}}=0
\ee
read $\phi_{\pm}$ 
\be
\phi_{\pm}=\frac{1}{2}\left(-3\varepsilon\pm \sqrt{\Delta}\right)
\label{phiminasym}
\ee
where
\be
\Delta=4+9\varepsilon^2-4\rho
\label{deltamin}
\ee
In the case of explicit symmetry breaking (asymmetron wall formation) the symmetry gets restored, for a background density larger than the critical density corresponding to the symmetron field. For the asymmetron case we have 
$\rho>\rho_{*,as}=1+\frac{9}{4}\varepsilon^2$ which is larger than the critical density ($\rho_*=1$) in the symmetron case.

The form of the asymmetron effective potential in vacuum and in high density cosmological regions is shown in Fig. \ref{figpot}. Clearly, in the case of asymmetron model the two local minima of the potential depend on the matter density as in the symmetron case. In the presence of sufficiently high density, the symmetry is restored along with GR since the coupling $A(\phi)\approx1$. Thus a screened fifth force is associated with the  {\it asymmetron} field. However,  as indicated in Eqs. (\ref{phiminasym}), (\ref{deltamin}) and in Fig. \ref{figpot} in the case of asymmetron model the two local minima of the potential in low density regions are not symmetric $\phi_+\neq |\phi_-|$ and non-degenerate. Thus, since the degeneracy of the vacua is  broken, this double-well potential has a false vacuum and a true vacuum due to the explicit symmetry breaking induced by the cubic term. The difference between the false and true vacuum energies increases with $\varepsilon$ as 
\be
V_{eff}(\phi_+)-V_{eff}(\phi_-)=\left[2\varepsilon(1-\rho)+\frac{9}{2}\varepsilon^3\right]\sqrt{\Delta}
\ee
Clearly, the energy difference between the vacua increases linearly with $\varepsilon$ for small $\varepsilon$.

\subsection{Spherical wall interaction with a matter shell: A toy model}

We consider a  finite thickness spherical domain wall in the presence of spherical matter shell as a simple toy model (see Fig. \ref{figshells}). Although this model is too simple it enables us to draw useful conclusions. 

The scalar field energy of the system if the wall and the matter shells are separate  is approximated as\footnote{For simplicity, here we ignore the gradient energy which if included further enhances the attraction of the wall by the matter shell.}
\be
E_s=V(0)w_m R_m^2+V(0)w_w R_w^2
\ee
where $w_w$ ($R_w$) and $w_m$ ($R_m$) are the widths (radii) of the domain wall and the matter shells respectively. In the matter shell region the field is at the minimum of the effective potential ($\phi=0$) with energy density $\rho_\phi \simeq V(0)$ while at the domain wall radius the field is trapped at the local maximum of the effective potential ($\phi=0$) with the same energy density $\rho_\phi \simeq V(0)$.

The energy of the system if the wall and the matter shells overlap is 
\be
E_o=V(0)w_w R_w^2
\ee
where we assumed without loss of generality that $w_w>w_m$. Therefore, the energy difference of the two configurations is 
\be
\Delta E=E_o-E_s=-V(0)w_m R_m^2<0
\label{deso}
\ee
Thus $E_o<E_s$ and it is energetically favored for the wall to overlap with the matter shell. In contrast to the conventional domain walls, the symmetron and asymmetron walls tend to stay in regions where the matter density is high. This is confirmed numerically in what follows.

\begin{figure}
\centering
\includegraphics[width = \columnwidth]{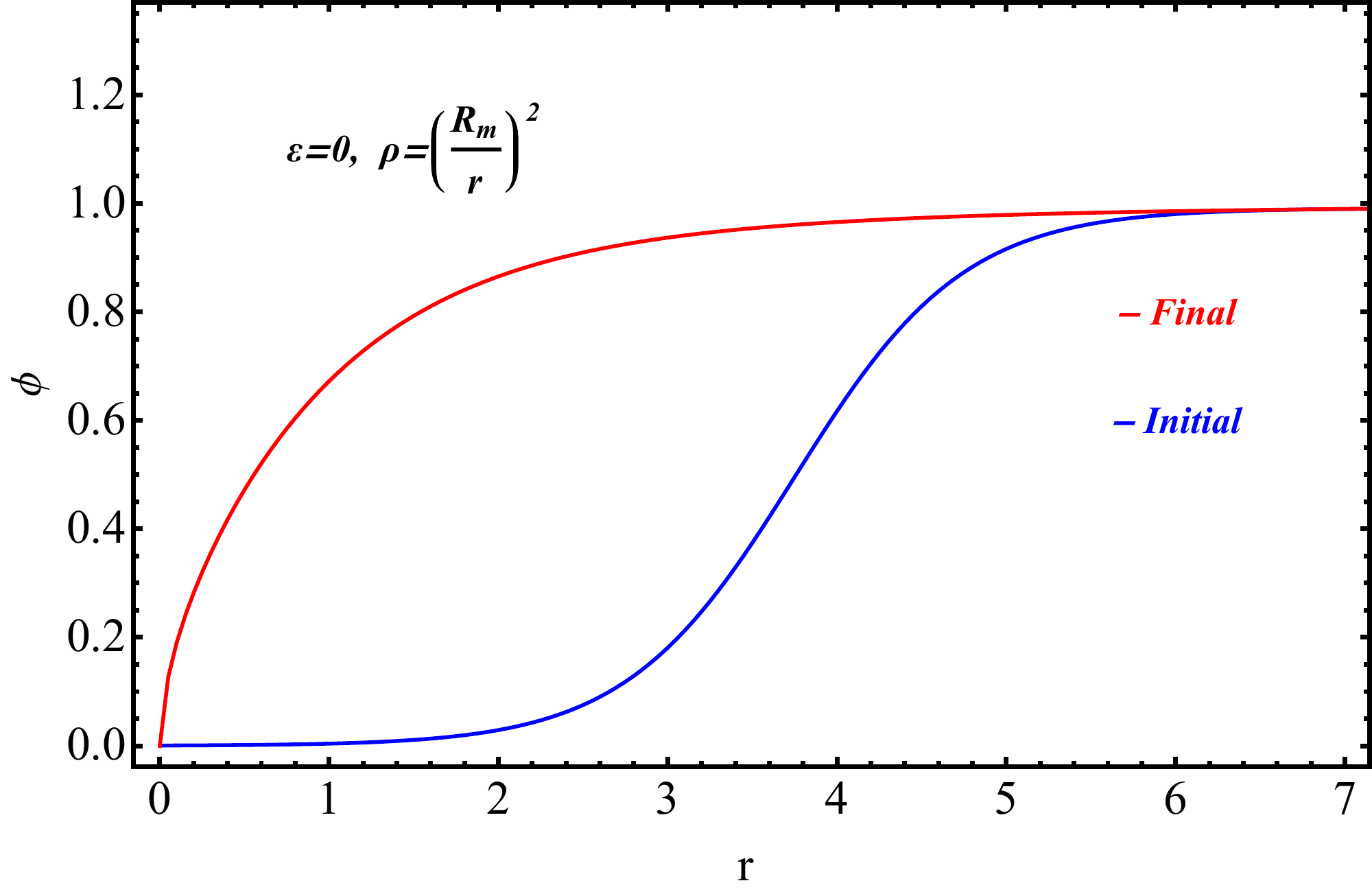}
\caption{The  scalar field $\phi$ as a function of the distance $r$ corresponds to the solution in the case of monotonic matter density increasing towards the center (with $R_m=1$). In this resulting minimum energy field configuration we see  a collapse of the wall due to tension. The energy minimization was performed numerically using $N=150$ lattice points.}
\label{figsmon}
\end{figure} 

\section{Static stable spherical wall configurations in the presence of matter} 
\label{staticsol}

\subsection{Analytic considerations}
\label{Analytic considerations}

A spherical domain wall is a field configuration that interpolates between the two minima $\phi_{\pm}$ of the effective potential as the surface of the wall sphere in physical space is crossed. The wall is characterized by the surface energy density $\sigma$  \cite{Vachaspati:2006zz} which depends not only on the configuration of $\phi$ but also on the matter density $\rho$. The corresponding to tension force per unit area is $p_{\sigma}\sim \sigma/R(t)$ (with $R(t)$ the curvature scale). In addition a pressure difference $p$ (with $p\sim \varepsilon \eta^3$ for $\rho\ll\rho_*$ \cite{Vachaspati:2006zz}) pushes the wall toward the vacuum with the lowest energy (true vacuum). 
The dynamics of the spherical asymmetron wall surrounding a true vacuum region is determined by three factors:
\begin{itemize}
    \item The tension term that favors contraction of the spherical wall with contribution to the energy $E_\sigma \sim \eta ^3 R(t)^2  $. This energy term increases with the wall radius.  
    \item The vacuum energy difference term that favors expansion of the true vacuum domain with contribution to the energy (relative to the exterior false vacuum domain) $E_{vac}\sim -\varepsilon \eta^3 R(t)^3$ for small $\varepsilon$. This negative energy term decreases with wall radius $R(t)$ and favors expansion. If the wall surrounds a false instead of a true vacuum region, then the sign of $E_{vac}$ will be positive  and the fate of the wall radius in the absence of the coupling to matter is contraction and collapse due to both tension and false vacuum energy.
    \item The term due to the coupling to matter $E_{mat}\sim -w_m R_m^2 \eta^4 \delta(R(t)-R_m)$ which dominates over the effect of tension as shown in Eq. (\ref{deso}) when the wall overlaps with the matter density shell. The $\delta$ function should be replaced by a smooth function leading to an attractive force, in thick-smooth realistic density profiles as those discussed in the next section.
\end{itemize}
The first two terms can at best lead to an unstable spherical domain wall as it can easily be verified that they lead to a static configuration at an energy maximum (instability) with respect to R. These configurations would tend to contract if the initial spherical wall radius is less than a critical value and would tend to expand if the initial radius is larger than this value. This could have been anticipated also due to Derrick's theorem \cite{Derrick:1964ww}. Stability can only be achieved due to the last term which is due to the external coupling to the matter density shell which violates the assumptions of Derrick's theorem and allows for a stable static spherical wall configuration as demonstrated numerically in what follows. A similar stabilization mechanism has been recently considered using external gravitational fields instead of a coupling to matter density \cite{Alestas:2019wtw}.

\subsection{Numerical energy minimization}
The evolution of the spherical domain wall is described by the action (\ref{action}) and the corresponding dynamical equation (\ref{eomres}). The energy of the spherical wall, assumed initially static is given by Eq. (\ref{energyres}).
 
We search for a stable static wall configuration by minimizing the discretized integral of the field energy Eq. (\ref{energyres}) starting from an initial guess that interpolates between the two vacua $\phi_+$, $\phi_-$ at a radius $R_w$
\be 
\phi(r)=\frac{\phi_+ -\phi_-}{2}\;\tanh(\frac{r-R_w}{w_w})+\frac{\phi_+ +\phi_-}{2}
\label{inguess}
\ee
We have verified that the precise form of the initial guess does not affect the final field configuration that minimizes the energy.

The boundary conditions may be set such that the spatial $r$ derivative of the scalar field is 0 at the two boundaries of $r$ ($r=0$ and $r=r_{max}$). Alternatively the boundary condition can fix the field at the corresponding vacua at the two boundaries. Both types of boundary conditions lead to the same minimum energy static field configuration in the cases studied.

Therefore, a simple way to derive numerically the basic features of the evolution of the wall initial configuration Eq. (\ref{inguess}) is to explicitly minimise the energy functional Eq. (\ref{energyres}) with fixed boundary conditions. We thus use the
Energy Minimization (EM) method which consists of the following steps:
\begin{enumerate}
    \item 
We discretize the energy functional Eq. (\ref{energyres}) as a sum over $N$ lattice points as
\be
E=dx\sum_{i=1}^N\left[r_i^2\left(\frac{\phi_i-\phi_{i-1}}{dx} \right)^2+r_i^2V(\phi_i) \right]
\label{energydis}
\ee
where $r_i=idx$, $dx=r_{max}/N$ and $\phi_i\equiv\phi(r_i)$.
    \item
We numerically minimize the sum (\ref{energydis}) with respect to the N lattice values of the field $\phi_i$ (one value at each lattice point) keeping fixed the boundary conditions.
\end{enumerate}
\begin{figure}
\centering
\includegraphics[width = \columnwidth]{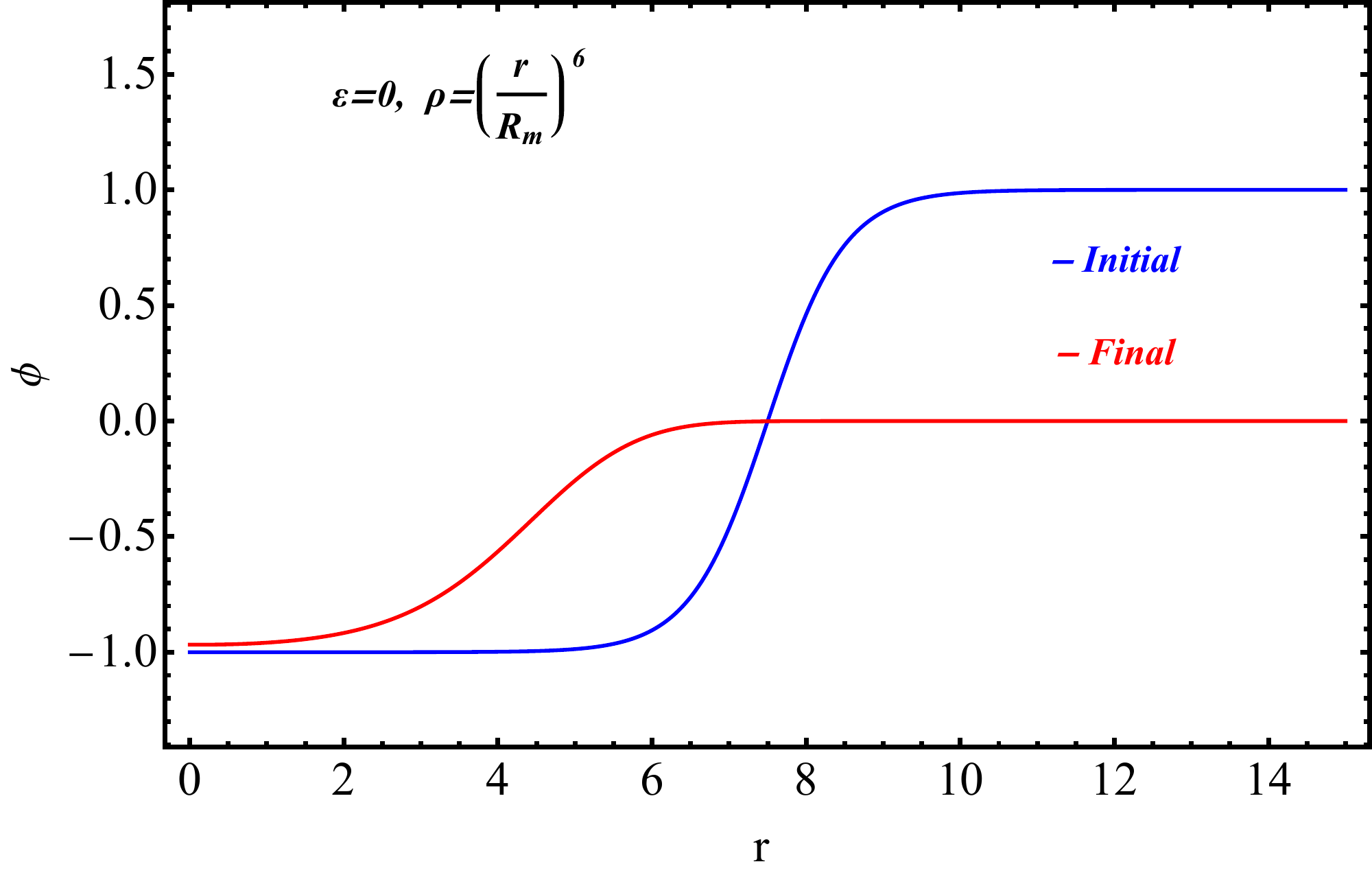}
\caption{The  scalar field $\phi$ as a function of the distance $r$ corresponds to the solution obtained from the energy minimization method in the case of increasing matter density. This field configuration appears to be stabilized by the combined effects of the wall tension and the attraction of the increased matter density as $r$ increases.}
\label{figs6}
\end{figure} 

\begin{figure*}
\centering
\includegraphics[width = 0.96\textwidth]{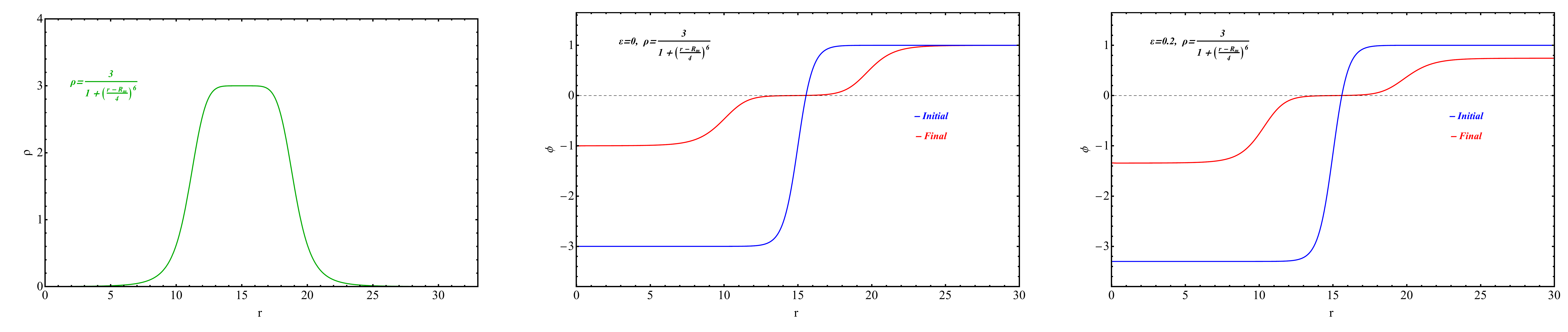}
\caption{Left panel: The matter density of the spherical matter shell of the form  (\ref{matshell}) with radius $R_m = 15$. Middle panel: The scalar field $\phi$ in symmetron case ($\varepsilon=0$) as a function of the distance $r$ corresponds to the solution obtained from the energy minimization method in the case of matter density of the spherical matter shell of the form  (\ref{matshell}) with radius $R_m=15$. The final minimum energy configuration is independent of the initial guess shown here in blue. Right panel: The scalar field $\phi$ in asymmetron case ($\varepsilon=0.2$) as a function of the distance $r$ corresponds to the solution obtained from the energy minimization method in the case of matter density of the spherical matter shell of the form  (\ref{matshell}) with radius $R_m=15$. }
\label{figshellall}
\end{figure*}

\begin{figure*}
\begin{centering}
\includegraphics[width=0.96\textwidth]{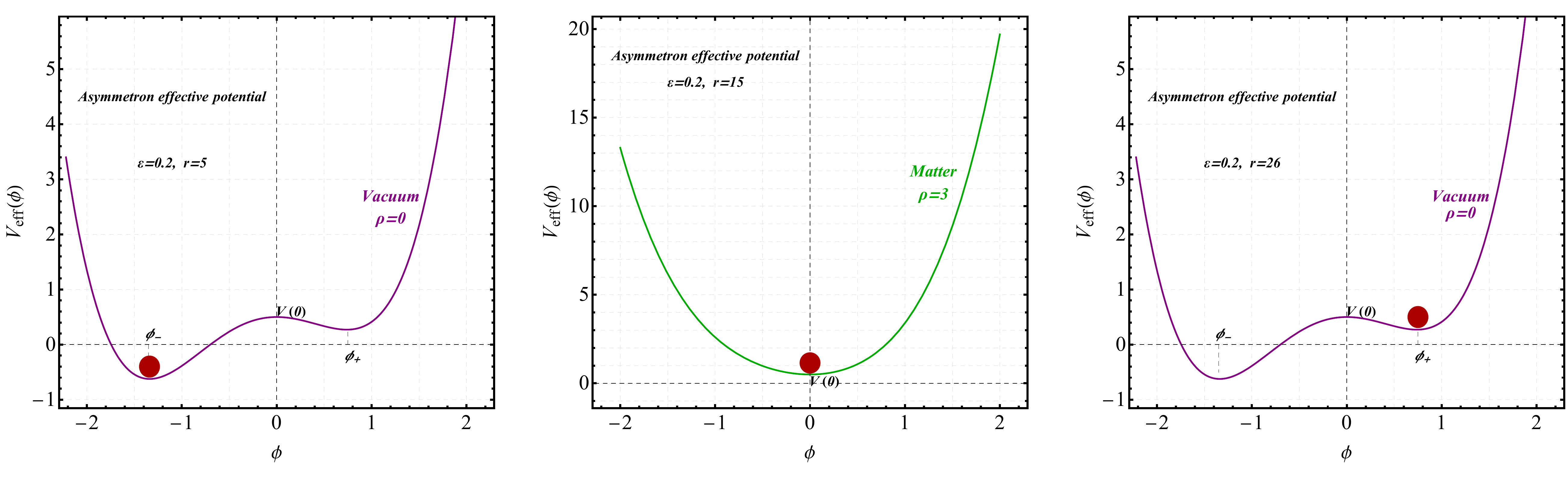}
\par\end{centering}
\caption{The form of the asymmetron (with $\varepsilon=0.2$) effective potential for the case $\rho=0$ (vacuum) and $\rho=3$ (high density) (see Figs. \ref{figpot} and \ref{figshellall}). The red points represent the value of the field and show how the asymmetron field changes as the wall is crossed by increasing $r$.} 
\label{figr} 
\end{figure*}

\begin{figure}
\centering
\includegraphics[width = \columnwidth]{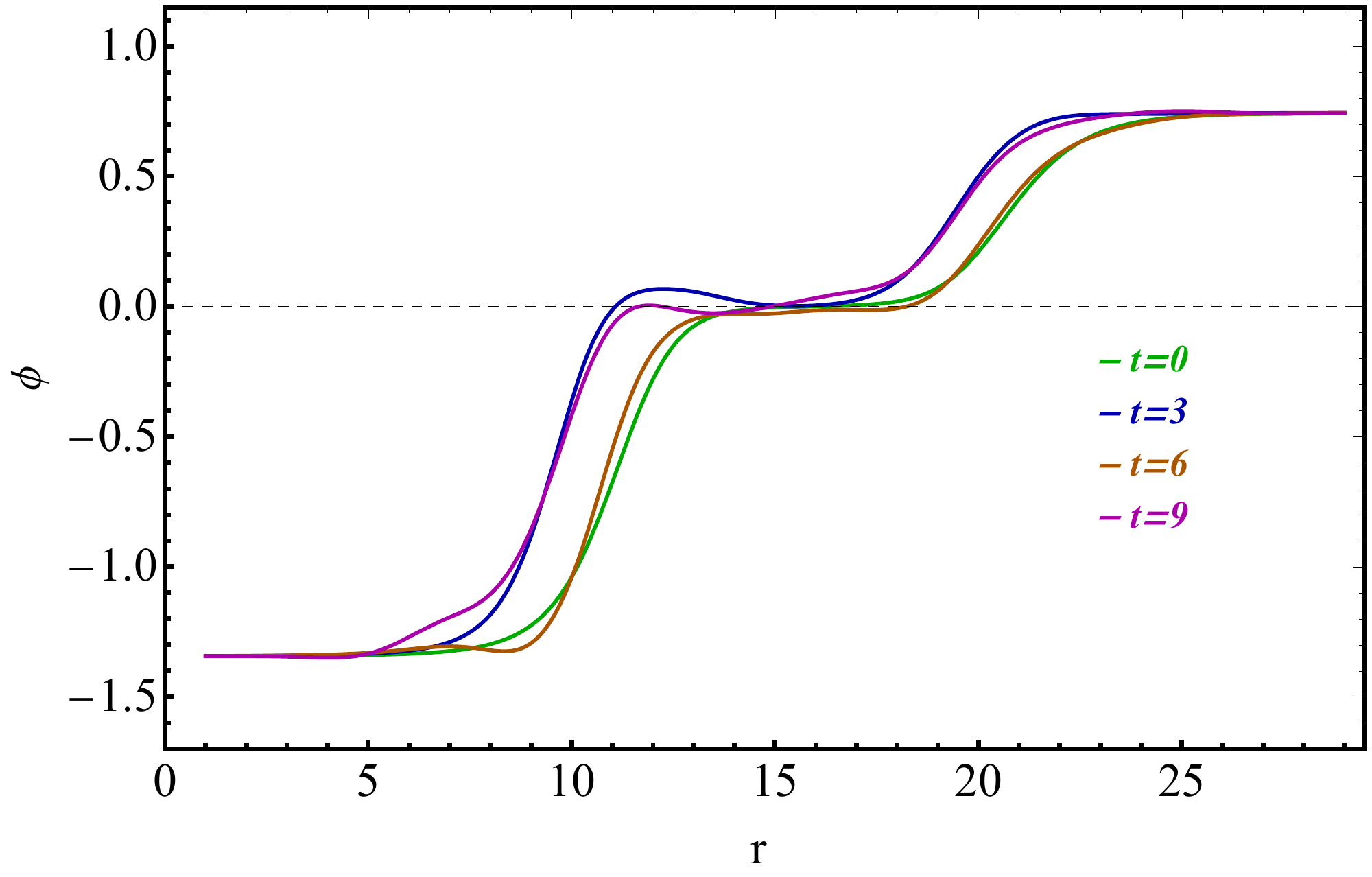}
\caption{Simulation of time
evolution of the perturbed scalar field corresponding to a perturbed spherical asymmetron domain wall. The wall gets trapped at the matter shell as expected (collapse is avoided).} 
\label{figpet} 
\end{figure}

In particular, we consider the following cases:\\

{\bf I. Spherical Symmetron Walls}\\

We first allow only spontaneous symmetry breaking and set $\varepsilon=0$. We consider the following matter density profiles:
\begin{itemize}
\item
{\bf Monotonic matter density increasing towards the center} of the form:
\be
\rho(r)=\left(\frac{r}{R_m}\right)^{-2}
\ee
with $R_m=1$.

We fix the boundary conditions such that the field remains at the corresponding vacuum on each boundary (we choose $\phi_+$ at the outer boundary where the matter density is low).
\be
\phi(r=0)=0, \,\,\phi(r=r_{max})=\phi_+
\ee
According to the above analytic arguments we anticipate an attractive force of the wall towards the center where the matter density is maximum in addition to the tension force which further amplifies this trend for collapse. In Fig. \ref{figsmon} we show the initial guess wall configuration and the final configuration emerging after the EM method. The minimization of the energy leads to a collapse of the wall due to tension  as expected.

\item
{\bf Increasing outward matter density} of the form:
\be
\rho(r)=\left(\frac{r}{R_m}\right)^6
\ee
with  $R_m=5$ and the following two boundary conditions
\be
\phi'(r=0)=0, \,\,\phi'(r=r_{max})=0
\ee
Unlike the result of the previous case, here we anticipate an outward force driving the wall radius to larger values where the density is larger. This trend is expected to compete with the wall tension. Indeed, here the field configuration emerging after EM method appears to be stabilized by the combined effects of the wall tension and the attraction of the increased matter density as $r$ increases. This resulting field configuration is shown in  Fig. \ref{figs6}.

\item 
{\bf Shell-like matter density} of the form (see in left panel of Fig. \ref{figshellall})
\be
\rho(r)=\frac{3}{1+\left(\frac{r-R_m}{4}\right)^6}
\label{matshell}
\ee
with $R_m=15$ and boundary conditions:
\be
\phi'(r=0)=0, \,\,\phi'(r=r_{max})=0
\label{bouncon}
\ee
As expected from the analytic arguments of Eq. (\ref{deso}) in this case the minimum energy field configuration corresponds to a wall radius overlapping with the matter shell radius (see in middle panel of Fig. \ref{figshellall}). 
\end{itemize}

{\bf II.  Stable Spherical Asymmetron Walls}\\

In the presence of an explicit symmetry breaking leading to the asymmetron field, the above results remain qualitatively unaffected. In this case we set $\varepsilon=0.2$  and assume a shell-like spherical matter density of the form  (\ref{matshell}) (see in left panel of Fig. \ref{figshellall}) and boundary conditions (\ref{bouncon}).

The resulting asymmetron field configuration after energy minimization is shown in right panel of Fig. \ref{figshellall}. The form of the corresponding asymmetron effective potential and the field values as the distance from the center of the spherical matter overdensity increases is shown in Fig. \ref{figr}. Snapshots of the potential and the corresponding field values are shown for matter density $\rho=0$ (vacuum inside and outside the matter shell) and $\rho=3$ (on the matter shell).   The red points represent the position of the field and show how the field changes as the distance $r$ from the center increases.

In order to further confirm the stability of the derived minimum energy configurations $\phi_s(r)$ we have perturbed them and implemented numerical dynamical evolution using a explicit Runge–Kutta algorithm \cite{10.5555/1403886}. In particular, we solve numerically Eq. (\ref{eomres}) with initial conditions
\be
\phi(0,r)=\phi_s(r-\delta  r),\,\,\,\dot{\phi}(0,r)=0
\ee
with  boundary conditions
\be
\phi(t,0)=\phi_s(0),\, \,\,\dot{\phi}(t,0)=0
\ee 
\be
\phi'(t,r_{max})=0, \,\,\dot{\phi}(t,r_{max})=0
\ee 
where the dot denotes differentiation with respect to the cosmic time $t$ and prime denotes differentiation with respect to the distance $r$.

The imposed perturbations on the minimum energy configuration of the right panel of Fig. \ref{figshellall} correspond to an initial shift by $\delta r <1$ of the wall radius $r$. 

The evolved scalar field configuration corresponds to a spherical wall with a radius that appears to be oscillating around the radius of the matter shell, effectively being trapped by it as shown in Fig. \ref{figpet}. This behavior is consistent with the stability of the spherical wall implied by both the analytic arguments of Subsection \ref{Analytic considerations} and by the energy minimization procedure discussed above.

\begin{figure*}
\centering
\includegraphics[width = 1\textwidth]{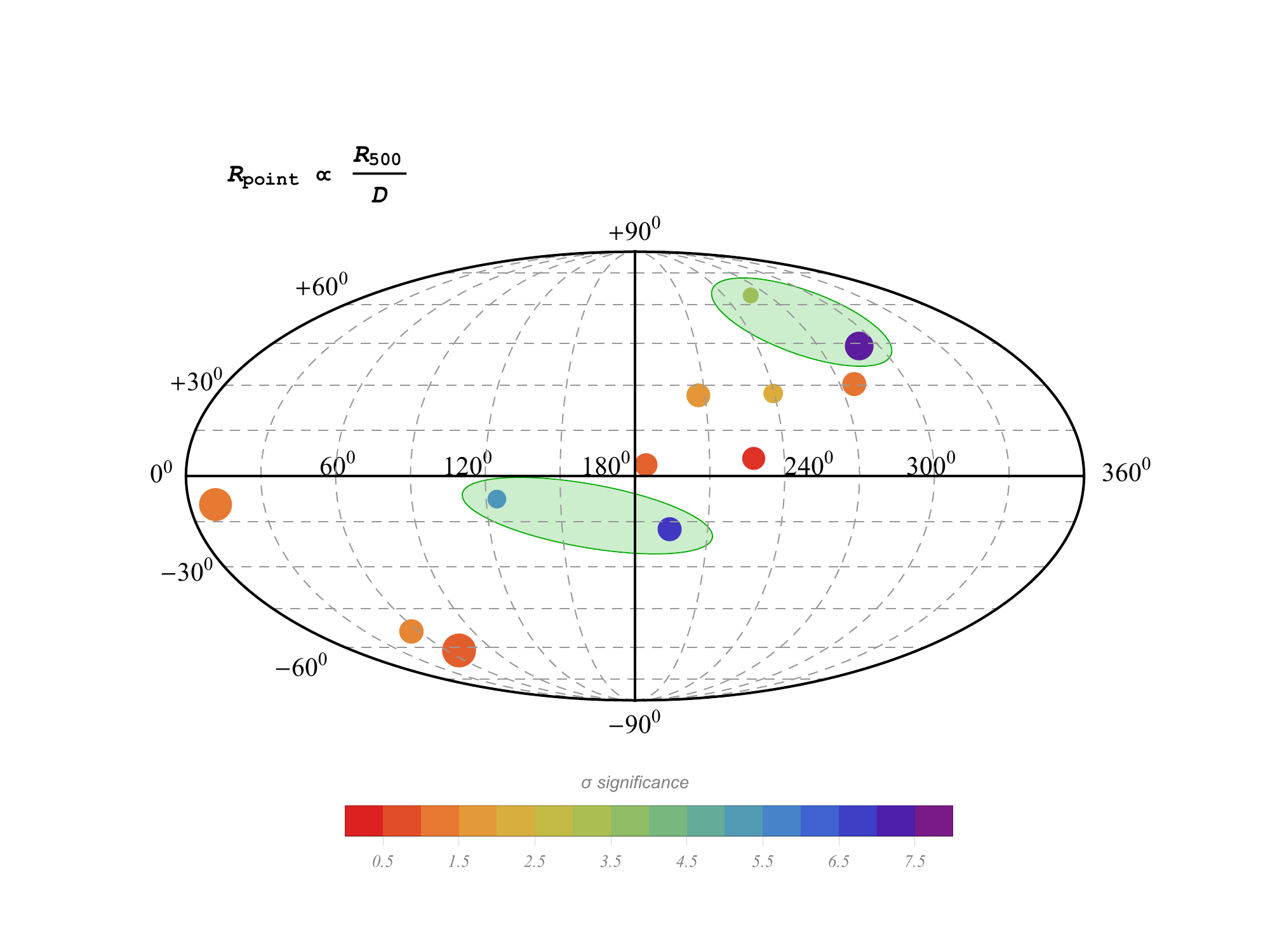}
\caption{Mollweide projection view of 12 cluster locations of Ref. \cite{Haridasu:2021hzq} in galactic coordinates (see Table \ref{tab:cluster}). The colour of the points on the plot corresponds to their $\sigma$ significance for a deviation from the GR, which is indicated in the horizontal colour bar. Four clusters  (in shaded green regions-bubbles) have large negative value for $\Xi_1$ parameter, significantly ($>3\sigma$) different from the GR ($\Xi_1=0$) expectation. The size of the points $R_{point}$ was designed according to the size of the clusters $R_{500}$ and their distance $D$.} 
\label{figcluster} 
\end{figure*}



\subsection{Observational considerations}

If asymmetron walls exist in Nature there could be cosmological regions bounded by surface-like matter overdensities where the strength of gravity would be different from other regions. Thus, the expansion rate within these regions would be different as would be the growth rate of cosmological perturbations and formation of structure. This inhomogeneity of the expansion rate could be detectable as anisotropies of the SnIa luminosity distances at a given redshift and could also be related with some of the observed cosmic dipoles (alpha dipole, quasar dipole etc). These observations could be used to impose bounds on the explicit symmetry breaking parameter $\varepsilon$.

The variation of the growth rate of cosmological perturbations among different domains could manifest itself as variation of the cluster properties including the cluster pressure and density profiles \cite{1989ApJ...341L..71E,1989ApJ...347..563P} (see for a review in Refs. \cite{Allen:2011zs,Kravtsov:2012zs}). Such variation in cluster properties which could be associated with properties of gravity has recently been identified in Ref. \cite{Haridasu:2021hzq}. In what follows we explore the possible relevance of the results of Ref. \cite{Haridasu:2021hzq} with the existence of asymmetron walls and their corresponding prediction for the existence of spatial cosmological domains with distinct properties of gravity.

Galaxy clusters are the largest gravitationally bound structures of the Cosmic Web. Thanks to the various surveys using dynamical, kinematic and weak lensing tracers, galaxy clusters can be used as a powerful cosmological probe of gravitational theories \cite{Sakstein:2016ggl,Salzano:2017qac,Haridasu:2021hzq,Laudato:2021mnm} and screening mechanisms \cite{Terukina:2013eqa,Wilcox:2015kna,Pizzuti:2020tdl}. 

The cluster pressure and density profiles can be inferred using the Sunyaev-Zeldovich effect, the inter galactic gas, the so called Intra Cluster medium (ICM), the temperature from their X-ray emission and the velocities of the individual cluster members.  These profiles can be used to search for possible changes of properties of gravity in different domains of cosmological space.

Recently the authors of Ref. \cite{Haridasu:2021hzq} have used cluster profile properties to test and constrain the parameters of the Degenerate Higher-Order Scalar-Tensor (DHOST) theory \cite{Langlois:2015cwa} (see also Refs. \cite{Zumalacarregui:2013pma,Langlois:2015skt,Crisostomi:2016czh,BenAchour:2016cay,Motohashi:2016ftl,BenAchour:2016fzp,Langlois:2017mxy} for recent related studies and Refs. \cite{Langlois:2018dxi,Kobayashi:2019hrl} for relevant reviews).  


The modified gravitational potential for the DHOST theory in the galaxy cluster as a static spherically symmetric object is \cite{Kobayashi:2014ida,Crisostomi:2017lbg,Langlois:2017dyl,Bartolo:2017ibw,Dima:2017pwp}
\be
\frac{d\Phi(r)}{dr}=\frac{G_{eff}M(<r)}{r^2}+\Xi_1 G_{eff}\frac{d^2M(<r)}{dr^2}
\ee
where $M(<r)=\int_0^r 4\pi r'^2\rho(r')dr'$ is the total mass (dark matter, gas, and galaxies) within the radial distance r, $G_{eff}=\tilde{\gamma}G$ is the effective Newton’s constant 
and $\Xi_1$ is a dimensionless parameter which depends on the noniminal coupling of the DHOST theory.
The modified gravity parameters $\tilde{\gamma}$ and $\Xi_1$ can be recognized as quantifying the deviation of the DHOST theory from GR, which is recovered for $\tilde{\gamma}=1$ and $\Xi_1=0$.

Ref. \cite{Haridasu:2021hzq} uses cluster data profiles of the XMM-Cluster Outskirts Project (X-COP) \cite{Eckert:2016bfe} to  place constraints on the DHOST parameters defining the deviation from GR. This very large programme uses a joint analysis of XMM-Newton and Planck data and targets the outer regions ($R>R_{500}$\footnote{For a given overdensity $\Delta$, the radius $R_{\Delta}$ is determined as the distance from the halo centre within which the mean density is $\Delta$ times the critical density, $\rho_c(z)=3H^2(z)/(8\pi G)$, at the halo redshift. Thus $\Delta \rho_c(z)=M_{\Delta}/(4/3\pi R_{\Delta}^3)$, where $M_{\Delta}$ is the halo mass i.e. the mass enclosed in $R_{\Delta}$.}) of a sample
of 13 massive  ($10^{14}M_{\astrosun} \lesssim M_{500}\lesssim 10^{15}M_{\astrosun}$)  local galaxy clusters in the redshift range $0.04<z<0.1$ at uniform depth. 

The constraints on the DHOST parameter $\bar{\gamma}=\tilde{\gamma}\times M_{500}/M_{500}^{GR}$ and $\Xi_1$ as obtained for each of the clusters by Ref. \cite{Haridasu:2021hzq} and the corresponding $\sigma$ significance for deviation from GR expectation are shown in Table \ref{tab:cluster} of the Appendix \ref{sec:Appendix_A}. As illustrated in Fig. \ref{figcluster} 4 clusters (A644, A1644, A2319 and A2255) have large negative value for $\Xi_1$ parameter, significantly ($>3\sigma$) different from the GR. Also for these 4 clusters we have $\bar{\gamma}<1$ ($G_{eff}<G$) with $\sim 2\sigma$ significance for a deviation from GR. These cluster constraints may be either interpreted as upper bounds on deviations of the DHOST parameters from their GR values or in a less conservative approach as possible hinds for modification of gravity. On the contrary, the constraints on $\bar{\gamma}$  and $\Xi_1$  for the other 8 clusters are fully consistent with GR.

In Fig. \ref{figcluster} we show the green ellipses that surround observed regions-bubbles in space ($\sim 50 Mpc$) where clusters with hints of  weaker effective gravitational constant were found in Ref. \cite{Haridasu:2021hzq}. These spatial sectors where the properties of gravity may be distinct from other regions may be consistent with the existence of asymmetron walls separating these sectors from other spatial sectors with slightly different properties of gravity. \\

\section{CONCLUSION-DISCUSSION}
\label{CONCLUSION-DISCUSSION}

We have generalized the symmetron screening mechanism by allowing for an explicit symmetry breaking of the symmetron $\phi^4$ potential by the cubic term $\varepsilon \phi^3$. In such a screening scalar field (the {\it  asymmetron})  the two local minima of the potential in low density regions are neither degenerate nor symmetric ($\phi_+\neq -\phi_-$). Thus the asymmetron domain wall network that may form includes a transition in the value of the Jordan frame effective gravitational constant as the asymmetron wall is crossed. 

We have  implemented numerical energy minimization and simulation of  evolution of  spherical symmetron and asymmetron domain walls in the presence of a matter shell. We have thus demonstrated that the walls get trapped by  matter overdensity shells as expected preventing the collapse of spherical symmetron and asymmetron walls and leading to stable spherical wall configurations. We have used a simple analytical energetic argument to describe this stabilization mechanism. The relevance of these asymmetron wall configurations with recent cluster profile data which may be interpreted as hinting towards distinct gravitational properties of certain clusters has also been discussed.

The possible existence of an asymmeron wall network pinned on matter overdensities separating regions with distinct gravitational properties could constitute a physical mechanism for the realization of gravitational transitions in redshift space that could help in the resolution of the Hubble and growth tensions as described in the Introduction. In this context, a wide range of possible extensions of the present analysis could be considered. Such extensions include the following:
\begin{itemize}
\item
The search for anisotropies of the Hubble expansion rate in certain cosmological regions surrounded by matter overdensities which can not be explained by the observed sign and level of matter underdensities. If such local modifications of the Hubble expansion rate can not be explained by matter underdensities, they could be attributed to local modifications of the Friedmann equation due to local modifications of the properties of gravity.
\item
The comparison of the growth rate of cosmological perturbations in different cosmological spatial sectors using for example weak lensing, cluster count and/or redshift space distortion data.
\item
The implementation of N-body simulations is order to identify signatures of asymmetron walls on the large scale structure power spectrum and on the ISW effect.  
\item
The construction of other physically motivated mechanisms that could lead to spatial gravitational transitions at low redshifts e.g. in the context of scalar tensor theories false vacuum decay.
\end{itemize}

In conclusion, the asymmetron model offers an interesting novel approach for the modification of GR in distinct spatial sectors. The predicted gravitational transition in redshift space could lead to the resolution of the important cosmological tensions of the standard $\Lambda$CDM cosmology  \cite{Abdalla:2022yfr,Perivolaropoulos:2021jda,DiValentino:2021izs,Verde:2019ivm}. Observable new effects and new physics beyond the standard model could also be realized in the context of the asymmetron domain wall network and corresponding constraints on the explicit symmetry breaking parameter can be imposed.\\

\section*{Numerical Analysis Files}

The numerical files for the reproduction of the figures can be found in the \href{https://github.com/FOTEINISKARA/Gravitational_transitions_via_the_explicitly_broken_symmetron_screening_mechanism}{Gravitational transitions via the explicitly broken symmetron screening mechanism} Github repository under the MIT license.\\

\section*{Acknowledgments}
We thank David Mota, Øyvind Christiansen and Mona Jalilvand for their useful comments at the early stages of this project. This project was supported by the Hellenic Foundation for Research and Innovation (H.F.R.I.), under the "First call for H.F.R.I. Research Projects to support Faculty members and Researchers and the procurement of high-cost research
equipment Grant" (Project Number: 789).\\

\appendix
\section{Cluster collection} 
\label{sec:Appendix_A}

In this appendix we present the collection of 12 clusters. 

\centering
\begin{table*}

\begin{center}
\caption{The collection of 12 clusters. From left to right the columns correspond to: Abell names, galactic coordinates (from NED), redshifts (from NED), luminosity distances (from NED), the halo radii for overdensity of $\Delta=500$ with respect to the critical density of the universe at the cluster's redshift, the  modified gravity parameters $\Xi_1$ and $\bar{\gamma}$ which track the departure of DHOST theory from GR as derived by Ref. \cite{Haridasu:2021hzq} and the corresponding $\sigma$ significances.} 
\label{tab:cluster}
\begin{tabular}{ | c | c | c | c | c | c | c | c | c|c|c| }

\hline
 &&&&&GR &DOST&&&&\\
Cluster & RA & DEC  & $z$ &  D & $R_{500}$ & $R_{500}$ & $\Xi_1$&Significance&$\bar{\gamma}$&Significance \\
& $[Deg]$ & $[Deg]$ &  &  $[Mpc]$ &  $[Mpc]$ &$[Mpc]$ & & $\sigma_{\Xi_1}$& &$\sigma_{\bar{\gamma}}$\\ 
\hline 
\hline
 &&&&&&&&&&\\
A85 &10.458750&-9.301944&0.05506 &248&$1.270_{-0.015}^{+0.010}$&$1.292_{-0.030}^{+0.017}$&$0.30_{-0.27}^{+0.11}$&1.100&$1.05\pm 0.28$& 0.179\\
  &&&&&&&&&&\\
A644 &124.352083&-7.512778&0.07040 &332&$1.175_{-0.015}^{+0.020}$&$0.980_{-0.030}^{+0.028}$&$-1.04_{-0.19}^{+0.18}$&5.470&$0.58\pm 0.22$& 1.910\\
  &&&&&&&&&&\\
A1644&194.290417&-17.400278&0.04740 &222 &$1.003_{-0.017}^{+0.019}$&$0.844_{-0.027}^{+0.020}$&$-0.837_{-0.090}^{+0.119}$&7.034&$0.59\pm 0.16$&2.562\\
    &&&&&&&&&&\\
A1795&207.220833&26.595556&0.06248&293 &$1.150_{-0.010}^{+0.015}$&$1.101_{-0.035}^{+0.032}$&$-0.169_{-0.090}^{+0.111}$&1.523& $0.88\pm0.25$&0.480\\ 
 &&&&&&&&&&\\
A2029 &227.729167&5.720000&0.07872&372&$1.369_{-0.015}^{+0.019}$&$1.352_{-0.016}^{+0.089}$&$-0.04_{-0.12}^{+0.19}$&0.211&$1.03\pm 0.48$&0.063\\ 
 &&&&&&&&&&\\
A2142&239.585833&27.226944&0.09090&430&$1.389_{-0.017}^{+0.017}$&$1.326_{-0.024}^{+0.040}$&$-0.203_{-0.079}^{+0.101}$&2.010&$0.87\pm 0.38$&0.342\\ 
&&&&&&&&&&\\
A2255&258.129364&64.092572&0.08029&376&$1.180_{-0.021}^{+0.023}$&$0.953_{-0.043}^{+0.046}$&$-1.1_{-0.32}^{+0.26}$&3.438&$0.53\pm 0.28$&1.679\\ 
&&&&&&&&&&\\
A2319&290.286667&43.958333&0.05570&254&$1.336_{-0.006}^{+0.016}$&$1.151_{-0.016}^{+0.020}$&$-0.827_{-0.076}^{+0.108}$&7.657&$0.64\pm 0.15$&2.400\\ 
&&&&&&&&&&\\
A3158&55.724583&-53.635278&0.05917&273&$1.119_{-0.012}^{+0.016}$&$1.054_{-0.029}^{+0.057}$&$-0.23_{-0.18}^{+0.15}$&1.278&$0.83\pm 0.33$&0.515\\      
&&&&&&&&&&\\
A3266&67.850417&-61.443889&0.05906&273&$1.489_{-0.030}^{+0.027}$&$1.455_{-0.055}^{+0.045}$&$0.100_{-0.079}^{+0.137}$&0.730&$0.93\pm 0.65$&0.108\\          
&&&&&&&&&&\\
RXC1825&276.352917&30.441944&0.06500&299 &$1.108_{-0.012}^{+0.013}$&$1.130_{-0.018}^{+0.016}$&$0.17_{-0.13}^{+0.17}$&1.000&$1.06\pm 0.19$&0.316\\        
&&&&&&&&&&\\
ZW1215&184.419167&3.662500&0.07708&366&$1.368_{-0.029}^{+0.029}$&$1.331_{-0.076}^{+0.041}$&$-0.21_{-0.18}^{+0.27}$&0.778&$0.91\pm 0.64$&0.141\\     
&&&&&&&&&&\\                   
    
\hline
\end{tabular}
\end{center} 
\end{table*}

\raggedleft
\bibliography{Bibliography}

\end{document}